**Accepted manuscript.** This is the accepted version of an article forthcoming in *Philosophical Transactions of the Royal Society A*, as part of the special issue "New Approaches to the Foundations of Quantum Mechanics", edited by Philip Goyal, Jonte Hance, Harald Wiltsche, and Daniele Pizzocaro. The manuscript may be subject to changes during the journal's production process.

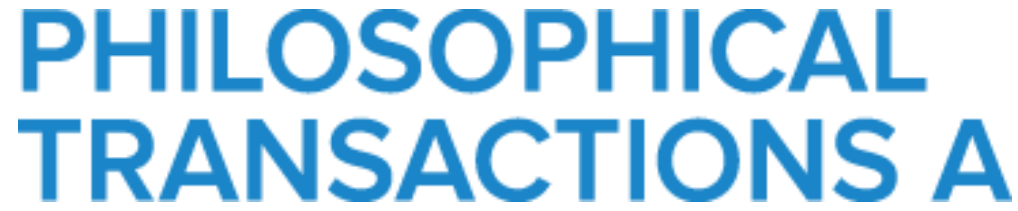

# Mutual Recognition in the Philosophy of Physics: QBism, Phenomenology, Hegel

**George Webster**[1]

[1] *Worcester College, University of Oxford; Worcester College, Walton Street, Oxford, OX1 2HB, United Kingdom;* *https://orcid.org/0000-0002-1591-8952*

---

Recent discussions of QBism have turned increasingly toward the question of its ontology. I argue that key insights into QBism's ontology of agency can be identified by reconstructing its response to Wigner's friend paradox. I clarify the relation between two formulations of that response: one grounded in the claim that quantum states are agents' personal probability assignments, and another grounded in the injunction to treat all users of quantum theory as agents on equal footing. Once disambiguated, these formulations reveal an implicit commitment to a mutually recognitive conception of agency, according to which the agential status of a user of quantum theory depends constitutively on the acknowledgement of other agents. This analysis clarifies what is at stake in recent QBist engagements with phenomenology—namely, the search for a more philosophically robust account of this conception of agency. I show how Merleau-Ponty's account of intersubjectivity underpins the recognitive structure implicated in QBism before tracing these themes to Hegel's account of mutual recognition. By locating QBism within a longstanding tradition of mutual recognition, I claim that we not only illuminate its theoretical structure but also open up novel conceptual resources for the rigorous articulation and justification of its ontology.

*Author for correspondence (george.webster@worc.ox.ac.uk).

†Present address: Worcester College, Walton Street, Oxford, OX1 2HB, United Kingdom

"For two users of quantum mechanics who interact, it requires each of them to treat their interaction as an action he or she takes freely on the other" [1, p1872].

"I am only genuinely free when the other is also free and is recognized by me as free" [2, §431A].

## 1. Introduction

Recent discussions of QBism have turned increasingly toward the question of its ontology. In this article, I focus on the ontology of agency and intersubjectivity implicated in the QBist response to Wigner's friend paradox. QBists often oscillate between two closely related formulations of their position: a formal response grounded in the claim that quantum states are agents' personal probability assignments, and a more explicitly agential response grounded in the need to treat all users of quantum theory as agents on equal footing—to "respect one's fellow" [1]. Yet the precise relation between these formulations is not always fully explicit. Part of the aim of this paper is to clarify this relation by reconstructing the conceptual connections between QBist notions of agency, measurement, and quantum-state assignment in a more systematic way.

This reconstruction, I argue, helps to clarify the conception of agency implicated in QBism's handling of intersubjective applications of quantum theory—one which, I suggest, is grounded in relations of mutual recognition. Though recent QBist appeals to phenomenology can be understood as attempts to furnish a more philosophically and explanatorily robust account of this conception, I argue that this recognitive structure cannot be understood apart from its origins in the nineteenth-century German idealism of G. W. F. Hegel. By tracing the provenance of the concept of mutual recognition, I aim to show how phenomenology and, more surprisingly, German idealism afford conceptual resources for the rigorous articulation and justification of QBist ontology.

The article proceeds as follows. Section 2 introduces the central tenets of QBism before reconstructing its response to Wigner's friend paradox. I disambiguate the relation between what I call the "standard" and "agential" formulations of their response, arguing that, once clarified, this response implicates a conception of agency grounded in relations of intersubjective recognition. Section 3 shows how phenomenology provides conceptual resources for a more philosophically robust account of this conception of agency, while Section 4 traces the relevant phenomenological themes back to Hegel's account of mutual recognition in his famous "lord-bondsman" dialectic.[1] I outline Hegel's logical account of the structure of self-consciousness, arguing that it affords an even stronger undergirding of QBism's conception of agency than that found in phenomenology. Section 5 concludes with some more general observations about the merits of and motivations for a pluralistic approach to the philosophy of physics.

## 2. QBism and the recognitive structure of agency

QBism is the claim that quantum mechanics describes normative constraints on its user's expectations—and those constraints alone. It arrives at this claim by embracing the "subjectivist" or "personalist" account of probability developed by de Finetti [3], Ramsey [4], and Savage [5], according to which probabilities express personal degrees of belief. The QBist applies this personalist account to the probabilistic content of quantum mechanics while also claiming that that content is exhaustive, since "[t]here is no mathematical fact embedded in a quantum state [...] that is not embedded in an appropriately chosen set of probabilities" [6, p391]. In other words, a quantum "state" plays no world-representing role, and this because it merely re-expresses one's personal probability assignments: "a quantum state is a personal judgment and not something mandated by the world external to the agent" [7, p87]. Accordingly, QBists make the Born rule the centrepiece of the quantum formalism, interpreting it as an addition to probability theory aimed at normatively guiding how one organises their degrees of credence [8]: "quantum mechanics is a tool anyone can use to evaluate [...] one's probabilistic expectations for one's subsequent experience" [9, p749].

A merit of this formal exercise is its neat resolution of the measurement problem. This problem concerns how we reconcile the linear dynamics of quantum mechanics with the non-linear "collapse" of the wavefunction when measured. Assuming that the formalism represents physical systems, one must deny at least one of the following claims: (i) that the formalism provides complete descriptions of quantum systems, (ii) that quantum systems always evolve according to the Schrödinger equation, or (iii) that measurements have unique determinate outcomes [10]. Bohmian mechanics denies (i), objective-collapse theories deny (ii), and the many-worlds interpretation denies (iii). The QBist, on the other hand, denies that the formalism represents physical systems. Since quantum states describe nothing more than my own expectations, what occurs when I make a measurement is no more mysterious than what occurs when I check my email inbox. I simply go from having some idea of what I *might* experience to having a more precise idea of what I *do* experience: "The notorious 'collapse of the wave-function' is nothing but the updating of an agent's state assignment on the basis of her experience" [9, p749], akin to having a suspicion confirmed or a wish fulfilled.

[1] Also translated as "master-slave" dialectic.

A natural rejoinder here is to enquire about the view's ontology. QBism may provide a coherent interpretation of the quantum formalism, one which neatly addresses various canonical paradoxes, but it must also provide some account of the world that explains why that interpretation is appropriate—that is, if the view is not a form of solipsism or instrumentalism, which QBists and (patient) critics alike agree it is not [11–17]. Though QBism's goal has always been "to say something deep about the character of reality", it has until recently maintained a deliberate and propaedeutic focus on more formal considerations—that is, on "get[ting] the "epistemics" of the theory right" [7, p78]. But now, having broadly assessed the technical implications of reconciling quantum theory with a personalist account of probability, QBists turn increasingly to the question of their ontology [7,18–22]. Here I focus on perhaps the most central dimension of that ontology—namely, the QBist conception of agents. I argue that the QBist response to Wigner's friend paradox reveals a formal account of agency based on relations of mutual recognition, and that phenomenology and German idealism afford conceptual resources particularly appropriate to clarifying and underwriting this account. Before this can be shown, however, an ambiguity between two formulations of the QBist response to Wigner's friend scenarios must be addressed.

The first formulation simply outlines the implications of QBism for intersubjective applications of quantum theory. In his famous thought experiment [23],[2] Wigner stands outside a sealed laboratory in which his friend performs a measurement on a quantum system—to determine the spin value of an electron along a particular axis, say. Wigner then performs his own indirect measurement by asking his friend about the result. According to the textbook presentation of quantum mechanics, a paradox emerges when we consider the interval between the friend's measurement and her reporting her result to Wigner. For Wigner, after all, his friend is part of that system, and so she herself is in a superposition of *having measured spin up* and *having measured spin down*. And yet, the friend obviously has a definite description in mind. We thus have two competing descriptions of the same quantum state: one (Wigner's) of indefinite superposition, and one (the friend's) of a determinate measurement outcome. The QBist responds not by adjudicating on whose description is correct, but rather, as with their response to the measurement problem, by denying that the formalism is in the business of representing the world in the first place. If quantum state assignments are merely expressions of one's degrees of belief in a given outcome, then no paradox arises. Wigner and his friend do not have in mind two irreconcilable descriptions of a single physical system. Rather, each has an independent account of their own individual experience: still anticipatory in the case of Wigner, and corroboratory in the case of his friend. The same line of reasoning is deployed against more recent and complex iterations of the thought experiment [1].

[2] In what follows, I cite a reprint of Wigner's original argument in [24].

This first formulation is often accompanied by another, according to which one resolves the paradox by taking seriously the agential status of all participants in Wigner-style scenarios. For example: "the paradoxical features [...] disappear once both friend and Wigner are understood as agents on an equal footing with regard to their individual uses of quantum theory" [1, p1859]; "the apparent contradiction [...] is due to a failure to treat one of the participants in the thought experiment as an agent in the full sense of the word" [1, p1861]; "The root of the paradox is the notion that Wigner's state assignment is able to capture everything that is relevant to the friend's experience" [20, p151]. This formulation relates to Wigner's own concern that his thought experiment encourages us "to deny the existence of the consciousness of a friend" [24, p177]. A more realist approach, for example, may attempt to resolve the paradox by treating one's friend as just another physical system, akin to "an atom which may or may not be excited" [24, p177]. Such an approach may resolve the paradox by making Wigner's description of events authoritative, but it does so at the cost of placing the friend in a state of "suspended animation" [24, p177], treating her experiential standpoint as exhaustively captured by Wigner's own state assignment.

Implicit in their presentation is the claim that these two formulations are somehow equivalent. *Prima facie*, however, this equivalence is not obvious, and this because mere recognition of agency seems insufficient for resolving the paradox. A realist, for example, can stipulate the existence of other autonomous users of quantum theory *and still run into the paradox*. This, after all, is how the paradox gains its force: it is a concern precisely because it spells trouble for our initial supposition that there are other agents. In order to make this equivalence explicit, the QBist must explain how respecting one's fellow amounts to the claim that quantum states are agents' personal probability assignments. I argue that we can render the QBist's logic more explicit by emphasising their injunction to take seriously each participant's status as an agent "in the full sense of the word" [1, p1861], where this qualifying phrase invites us to embrace the distinctive QBist account of agency and its attendant concepts.

This logic runs as follows. An *agent* is an entity with the capacity to act.[3] One type of action available to agents is *quantum measurement*,[4] which is just the sort of action for which our expectations regarding its outcomes are appropriately expressed in terms of the quantum formalism—that is, in terms of *quantum states*, which are an agent's personal probability assignments.[5] Thus, the agential formulation is equivalent to the standard formulation once we acknowledge that all the relevant concepts (agency, measurement, quantum state) are rooted in the practical activity of agents. A quantum state ascription just is an agent's personal probability assignment, and so acknowledgement of agency is acknowledgement of the possibility of such ascriptions thus conceived. Though one may worry that relying on *QBist* conceptions of the relevant concepts introduces circularity into this way of formulating their response, the aim is to show how QBism deals with the paradox. So, the assumption of QBism is a feature rather than a bug. The trick to understanding the equivalence of the standard and agential formulations is to be consistent in that assumption.

Having elucidated QBism's account of the intersubjective use of quantum theory, we can tease out its implications for a QBist conception of agency. First, though, a note about how we frame such intersubjective scenarios. Though we nominally acknowledge the perspectives of its characters, Wigner's friend paradox is typically presented from the privileged standpoint of a third-personal super-observer—i.e., *us*, the theoreticians whose job it is to adjudicate between the state assignments of Wigner and his friend. But this mode of presentation is itself at odds with QBism, according to which such assignments are ineliminably first-personal: "The very tone of the question eschews the QBist conception of quantum states as pluralistic and always tagged to a specific agent" [7, p. 121]. Accordingly, even our supposedly removed position as cogitators on the paradox is itself enmeshed within the scenario under consideration, since any putative super-observer's assessment will be just another personal judgement. The imperative to acknowledge agency thus applies not to some detached theoretical perspective, but to each individual user of quantum theory: "For two users of quantum mechanics who interact, [their reasoning consistently about one another] requires each of them to treat their interaction as an action he or she takes freely on the other" [1, p. 1872].

---

[3] "An agent is an entity that can freely take actions on parts of the world external to itself and for which the consequences of its actions matter to it" [7, p82].

[4] "A quantum measurement is an agent's action upon its external world. A quantum measurement is cut from the same cloth as any other action she might take upon her world, as for instance by crossing a street. What makes an action specifically "quantum" is when it is worth the agent's while to analyze her expectations for its outcomes in terms of the quantum formalism" [7, pp115–6].

[5] "A quantum state is an agent's personal judgment. It serves to tie together all her probability assignments for the outcomes of all measurement actions she might take upon a system" [7, p116].

It is here that the mutually recognitive structure of the QBist conception of agency becomes apparent. By interpreting quantum theory in terms of personal probability assignments, I limit the scope of my use of the theory and so respect my friend's status as a locus of action and conscious experience. At the same time, I protect myself against the agency-negating implications of the paradox. For if I adopt an interpretation that allows me to deny the consciousness of a friend, then I also sanction the same treatment of myself. Thus, it is not merely my consistent use of the theory that is at stake in my acknowledgement of other users of quantum mechanics, but also *my agency*—that is, my capacity to act and use quantum mechanics among others. The recognition of other agents is thus a constitutive feature of agency on the QBist account.

This is not to say that a QBist agent cannot represent other users of quantum theory *as though they are mere physical systems*. QBists admit explicitly that "each agent has a dual role as a physical system for the other agent" [1, p1872]. If we did not have such a role, then we could not form beliefs or expectations about others. The point, though, is that quantum state assignments *just are* those beliefs or expectations—that is, rather than descriptions of others themselves. And, recalling the equivalence of the standard and agential formulations discussed above, to conceptualise state assignments in this way just is to acknowledge the capacity of others to apply such assignments in intersubjective scenarios. The agent's dual role thus involves both hypothetical treatment as another physical system and genuine recognition as a user of quantum mechanics. Hence the accompanying QBist reference to "a quantum Copernican principle" [1, p1872] can be understood in both its original and a quasi-Kantian sense—that is, as simultaneously centring and de-centring the agent in complementary respects. My use of quantum theory conforms to my experience and expectations, but in recognising as much I concede that not all uses of quantum theory do the same: "Quantum theory is a single-user theory *for each of us*" [7, p116; my emphasis].

These observations place QBism within a rich tradition of theorising about mutual recognition. So far, however, the account we have reconstructed remains largely formal. QBism's response to Wigner's friend paradox shows that consistent intersubjective use of quantum theory requires agents to acknowledge one another as loci of action and experience, but it does not yet explain why agency should possess this recognitive structure in the first place, nor does it justify the non-solipsistic assumption that there are other users of quantum mechanics.[6] It is in this context that recent QBist engagements with phenomenology—and in particular the phenomenological ontology of Maurice Merleau-Ponty—can be understood. That said, these engagements remain somewhat general, citing "significant overlap between the project of finding [...] a QBist ontology and the philosophy of Maurice Merleau-Ponty and other phenomenologists" [20, p148], or claiming that Merleau-Ponty's phenomenology "begs to be compared to [...] Wigner and his friend" [7, p129]. In the next section, I show in more detail how phenomenology provides a more robust account of the intersubjective conditions of experience and thus a deeper philosophical underwriting of the QBist conception of agency.

---

[6] My claim is not that QBism is or entails solipsism; it is that the formal structure of QBism does not yet provide reasons to reject solipsism. In the next sections, I show how phenomenology and German idealism provide QBism with such reasons.

## 3. Phenomenology and the recognitive structure of experience

QBism requires users of quantum theory to acknowledge one another as autonomous agents. As Schack puts it, "QBism [...] leaves the fundamental autonomy of the other agent intact" [20, p150]. And yet, the QBist account remains largely formal, treating this autonomy as a basic commitment rather than explaining why agency should possess its mutually recognitive structure in the first place. Phenomenology, by contrast, seeks to uncover the invariant structures of lived experience and, in doing so, develops an account of agency according to which mutually recognitive relations between agents are implicated in experience itself. It thus affords a more robust explanatory underpinning for the conception of agency employed by QBism. In what follows, I reconstruct this phenomenological account of intersubjective recognition before turning to its significance for QBist ontology. I focus primarily on the work of Merleau-Ponty, partly because much engagement with QBism (and physics in general) by contemporary phenomenologists already focuses largely on Husserl [25–28],[7] and partly because I want to clarify what is at stake in QBists' own tentative engagements with phenomenology [7,19,20], which focus largely on Merleau-Ponty.

The characteristic claim of phenomenology is that lived experience constitutes the foundation for all knowledge. This is an anti-dogmatic claim, expressing a refusal to accept any belief without at least first tracing its origins in experience—understood as the ultimate grounds for justification.[8] To execute this tracing operation, the phenomenologist must first provide an account of the nature and structure of experience. And, crucially, they must do so without importing conceptual resources not licensed by that same experience. Accordingly, phenomenology requires commitment to a *descriptive* and *first-personal* methodology. The phenomenologist, in other words, must describe the most immediate and lived characteristics of experience—that is, *how experience is experienced*. Only then can we engage in sufficiently self-critical inquiry. As Merleau-Ponty puts it:

---

[7] An exception is [29].
[8] On the justificatory force of experience, see [27].

> All my knowledge of the world, even my scientific knowledge, is gained from my own particular point of view [...]. The whole universe of science is built upon the world as directly experienced, and if we want to subject science itself to rigorous scrutiny and arrive at a precise assessment of its meaning and scope, we must begin by reawakening the basic experience of the world of which science is the second-order expression. [30, pviii]

This is not to say that phenomenology is merely subjective or anecdotal. Rather, it is concerned with the essential features of experience as such. The paradigm case of such a feature is intentionality—that is, the *directedness* of experience toward objects: "Experiences, wishes, or desires [for example] are essentially characterized by their being directed at something *beyond themselves*" [31, p3]. The task of phenomenology is thus to identify and describe the different sorts of intentional structure (in perception, judgement, imagination, recollection, desire, etc.) and any other modes of experience. In doing so, it provides a rigorous analysis not only of different sorts of conscious state, but also of the framework of meaning and intelligibility that makes possible any objective inquiry or empirical science in the first place.

An upshot of this phenomenological stance is its treatment of scepticism about the external world. Such scepticism problematises our relation to the world by construing experience in advance as a collection of mental states or representations. In doing so, it situates each of us within what Merleau-Ponty calls an "impregnable subjectivity" [30, px], whose relation to an external world quite naturally demands explanation. Merleau-Ponty, however, regards this picture as "an incomplete form of reflection which loses sight of its own beginning" [30, px]. That is, the sceptical concern arises only because experience has already been conceptualised in terms not licensed by its immediate character. When we attend more carefully to experience as lived, we find that objects are not first encountered as inner representations from which we must infer an external reality, but rather simply as objects given in perception: "the world is always 'already there' before reflection begins—as an inalienable presence" [30, pvii]. Properly understood, experience thus reveals our relation to the world not as something requiring reconstruction from within subjectivity, but as a form of "direct and primitive contact" [30, pvii].

These observations reveal a subtle difference between QBism's response to the charge of solipsism and the corresponding phenomenological critique. Fuchs, for instance, argues that "[t]he entire basis for calling QBism solipsism is just short-circuited by the concepts QBism relies upon for its very starting point" [7, p92]. That is, QBism is definitionally incompatible with solipsism because it takes quantum measurement to be an action "on the *world external to the agent*" [7, p92]. The basic formal commitments of QBism thus exclude the possibility of solipsism. The phenomenologist, on the other hand, rejects solipsism not through sheer formal stipulation, but by exposing its dogmatic assumptions about the nature of experience. Phenomenology thus introduces a subtle explanatory twist. Rather than taking contact with the external world as a primitive assumption, it aims to show that attempts to deny such contact depend upon abstractions that are themselves philosophically contentious. We take ourselves to be in contact with the world not because we assume that we are, but because *we are too self-critical not to do so*.

A similar line of reasoning is deployed to account for one's relation to other agents—otherwise known as the problem of other minds. This problem can be expressed in terms of scepticism about a certain kind of analogical reasoning. The only mental states to which we have direct access are our own, and so we must make inferences about the mental states of others from associated physical interactions and bodily expressions. If I recoil and yelp after feeling the painful zap of a static electric shock, then I might infer that someone else experiences a similar sensation when I observe them reacting in the same way. Much ink has been spilled over concerns about the legitimacy of this sort of inference.[9] And yet, for Merleau-Ponty and other phenomenologists, the problem is ill-founded—and this because, as in our discussion above, it begins from a particular conception of human existence that fails to reflect experience. In particular, the sceptic "overestimates the difficulties involved in the experience of others and underestimates the difficulties involved in self-experience" [33, p182]. Our relation to ourselves is not merely one of transparent access to a private interior realm. Rather, much self-experience is mediated by the complexities of our embodied, social, and historical situation—a testament to the fact that, for Merleau-Ponty, my experience is "above all a relation to the world" [34, p117]. If this is correct, then the difference between self and other is less radical than the sceptic supposes. I encounter my own subjectivity not as an isolated consciousness hidden behind bodily appearances, but as already expressed in and through my worldly activity. Accordingly, the gestures, expressions, and conduct of others are not merely external signs from which hidden mental states must be inferred, but are rather among the ways in which subjectivity itself becomes manifest. As Merleau-Ponty puts it, "a perspective on the other is opened to me from the moment I define him and myself as "conducts" at work in the world" [34, p117].

This same thought is expressed more positively and in a more explicitly ontological idiom by Merleau-Ponty in his later work. He writes that any experiencing agent "must also be inscribed in the order of being" [35, p134] that they experience. To be a subject of experience, in other words, one must also be available to other experiencing subjects. Because one's perspective is itself situated within a world, it is necessarily accessible to other perspectives. In Merleau-Ponty's words:

[9] For a comprehensive historical survey of the problem of other minds, see [32].

> he who looks must not himself be foreign to the world that he looks at. As soon as I see, it is necessary that the vision [...] be doubled with a complementary vision or with another vision: myself seen from without, such as another would see me, installed in the midst of the visible, occupied in considering it from a certain spot. [35, p134]

Here the reciprocal structure of experience becomes apparent. My own situatedness within a world implies availability to other perspectives, just as the situatedness of others implies availability to mine. Experience is not achieved by an isolated subject; it occurs within a shared field in which subjects are mutually accessible to one another.

The point, however, is not merely that we are *accessible* to each other. If subjects are necessarily available to one another within the same field of experience, then they encounter one another not simply as observable objects but as fellow occupants of that field. The reciprocal perceptibility of subjects thus grounds a reciprocal acknowledgement of subjectivity itself: "my existence should never be reduced to my bare awareness of existing, [...] it should take in also the awareness that one may have of it" [30, ppxii–xiii]. The very structure of experience, in other words, constitutively involves recognition of other experiencers, and thus my very status as an experiencing subject depends upon relations of mutual recognition between subjects. Or, as Husserl puts it, "subjectivity is what it is—an ego functioning constitutively—only within intersubjectivity" [36, p172].

This recognitive structure is part of what makes phenomenology such a valuable resource for QBism. In our reconstruction in the previous section, we saw that QBism formally requires agents to acknowledge one another as autonomous centres of action and experience. The phenomenologist fleshes out this formal account by explaining not only how such acknowledgement is possible, but also that it is necessitated by a proper understanding of experience. Because subjects are reciprocally available to one another within a shared world, they encounter one another not merely as physical systems but as fellow experiencers. The autonomy of the other agent thus emerges not only as a formal stipulation but as a consequence of the structure of experience.[10]

---

[10] I should clarify the notions of "subject", "agent", and "autonomy", and how they relate, on my account. A subject is minimally autonomous in the sense that its experience is *its own*—that is, not constituted or determined externally, as in Wigner-style cases of suspended animation. Agency, as the capacity to act, involves a more substantial register of autonomy but nonetheless presupposes subjecthood. I do not mean here that agency requires full-blooded reflexive mental representations (e.g., intentions, beliefs, desires). But I do claim that agency depends upon an awareness of one's actions and their consequences to some at least minimally conscious or experiential degree. The QBist expresses this idea in an appropriately broad way by

In the following section, I trace the provenance of phenomenological accounts of intersubjectivity back to Hegel's famous account of mutual recognition. I then outline Hegel's own logical account before explicating its relevance to QBism.

## 4. Hegel and the recognitive logic of self-consciousness

It is difficult to overstate the influence of Hegel on twentieth-century phenomenology. His account of mutual recognition in particular is claimed to have had "an almost suffocating influence on left-wing existentialism" [38, p202].[11] This influence was mediated in large part through Alexandre Kojève's celebrated lectures [39], which played a central role in reintroducing Hegel to French intellectual life in the 1930s. Merleau-Ponty, who knew Kojève personally [40, p548], not only attended these lectures but reflects explicitly on the theme of mutual recognition throughout his writings. In his essay on "Hegel's Existentialism" [41], he writes:

> A more complete definition of what is called existentialism than we get from talking of anxiety and the contradictions of the human condition might be found in the idea of a universality which men affirm or imply by the mere fact of their being and at the very moment of their opposition to each other, in the idea of a reason immanent in unreason, of a freedom which comes into being in the act of accepting limits and to which the least perception, the slightest movement of the body, the smallest action, bear incontestable witness. [41, p70]

In this passage Merleau-Ponty contrasts his approach to other phenomenologists (e.g., Martin Heidegger, Jean-Paul Sartre, Simone de Beauvoir) who embrace "anxiety and the contradictions of the human condition" as characteristic themes of the phenomenological project. Instead, he argues that we better understand that project by appreciating its origins in distinctively Hegelian ideas—including "a universality which men affirm […] at the very moment of their opposition to each other", and "a freedom which comes into being in the act of accepting limits". As we will come to see in the discussion below, these are specific references to Hegel's account of mutual

---

stating that, for any agent, "its actions matter to it" [7, p82] (the use of "to" rather than "for" here is significant). Thus, to recognise a subject is to recognise a potential agent. For an overview of the philosophy of agency, see [37].

[11] For the purposes of this essay we can regard "existentialism" and "phenomenology" as interchangeable. Indeed, Merleau-Ponty identifies the two when he writes that "phenomenological reduction belongs to existential philosophy" [30, pxiv].

recognition. Thus, despite claims that this account "plays nowhere near as major a role in his thought as it does in the work of Kojève and Sartre" [38, p213], Merleau-Ponty in fact understands his philosophical project specifically in terms of mutual recognition.

Indeed, Merleau-Ponty deploys Hegelian resources to articulate his own account of intersubjectivity. Alluding to scepticism about other minds, and following a direct reference to Hegel's account of the life-and-death struggle, he writes: "For the struggle ever to begin, and for each consciousness to be capable of suspecting the alien presences which it negates, all must necessarily have some common ground and be mindful of their peaceful co-existence in the world of childhood" [30, p355]. Here Merleau-Ponty argues that such scepticism in fact presupposes and abstracts from a direct preliminary encounter with the other. And though he also draws on empirical evidence from child psychology, citing the implausibility of attributing complex analogical reasoning to young children who are nevertheless evidently responsive to the facial expressions of others,[12] he clearly regards this claim as a further expression and vindication of Hegel's account of a "struggle between consciousnesses, each one of which [...] seeks the death of the other" [30, p355]. Just as Hegel claims that "[s]elf-consciousness exists in and for itself when, and by the fact that, it so exists for another" [42, ¶178], so too does Merleau-Ponty insist that "we are collaborators for each other in consummate reciprocity" [30, p354]. The latter's account of intersubjectivity is thus deeply influenced by Hegel's own account of the structure of consciousness.[13]

Having motivated the move to Hegel historically, I turn now to motivate it conceptually. That is, in what follows I outline Hegel's logical account of mutual recognition, showing how it provides a suitable and particularly strong conceptual underpinning for a QBist ontology of agency. Before that, though, a contextual and methodological remark. Hegel's account of mutual recognition is set within his first and most famous major work, the *Phenomenology of Spirit*. Naturally, this is a highly dense and complex text, and it is beyond the scope of this essay to provide a full account of its project. That said, it will be helpful to note that "phenomenology" does not have the same meaning here as it does in the previous section.[14] Hegel is not interested primarily in our lived experience. Rather, the "experience" he seeks to relay is the perspective immanent to each meta-epistemological standpoint—that is, each candidate account of the relation of consciousness to its object—studied throughout the course of his text. In other words, Hegel is committed to a methodology of immanent critique. His aim is to expose an internal incoherence within a given standpoint and to follow its imagined efforts to iron out that inconsistency and to reaffirm its own self-conception. This observation should explain the somewhat anthropomorphised tone and quasi-mythic descriptive machinery used by Hegel (e.g., the language of desire, life-and-death struggle, and lordship and bondage). It would be a mistake, however, for the reader to infer imprecision or woolliness on Hegel's part. His account of the "experience" of certain shapes of consciousness may seem unorthodox, but it is a register through which he documents the logically necessary implications of those conceptual structures. As Houlgate observes: "he analyses the experience that, logically, consciousness *must* make—or *should* make—given its own particular certainties. The development of consciousness that Hegel describes in the *Phenomenology* is thus a necessary one" [46, p9]. This is the reason for my claim that phenomenology (in the sense of the previous section) and Hegelian philosophy occupy distinct explanatory registers and thus afford different kinds of justification for QBist ontology. Hegel's account of mutual recognition is logical rather than phenomenological. That is, instead of arguing that mutually recognitive relations between agents are implicated by the structure of experience, he contends that they obtain as a matter of logical necessity. Simply by reflecting on the conceptual structure of self-conscious agency, he claims, we are forced to understand it as grounded in relations of mutual recognition between multiple self-consciousnesses. With these clarifications in mind, we can turn now to consider Hegel's account.

[12] See also his essay "The Child's Relations with Others" in [34].
[13] For more on Hegel's influence on Merleau-Ponty, see [43,44].
[14] On this point, see J. N. Findley's forward to the *Phenomenology* [45, pxii].

Hegel starts his account of self-consciousness by observing an inconsistency in its most rudimentary form. On the one hand, self-consciousness is self-relating: it takes itself to be "a certainty which is identical with its truth" and thus "to itself its own object" [42, ¶166]. On the other hand, it is also non-reflexively conscious of an external world and so relates to something other than itself. Self-consciousness is thus caught in a contradiction: it seeks to understand itself as purely self-relating, yet can do so only in relation to an external object. Hegel argues that self-consciousness initially attempts to resolve this tension through what he calls "desire". Rather than treating the objects of experience as genuinely independent, it reconceptualises them as "only *appearance*" [42, ¶167]—that is, as phenomena whose significance derives from their relation to self-consciousness. It then enacts this reconceptualization through the activity of negation or consumption, demonstrating to itself that such objects possess no genuine independence but exist only as something available to it. In this way self-consciousness seeks to reaffirm its status as "only the 'I' itself" [42, ¶166].

And yet, this strategy proves to be self-defeating. The activity of negating otherness presupposes the presence of something other to negate and so remains constitutively dependent upon precisely what it seeks to eliminate. As Hegel puts it, "in order that this supersession can take place, there must be this other", and so "[d]esire and the self-certainty obtained in its gratification, are conditioned by the object" [42, ¶175]. The movement of desire therefore fails to secure the self-sufficiency sought by self-consciousness.

It is this failure of desire that leads self-consciousness to the revelation that it can achieve satisfaction "*only in another self-consciousness*" [42, ¶175]. Its attempt to recast itself as the activity of negating otherness is futile, since that activity itself relies constitutively on otherness. Self-consciousness must therefore somehow affirm its self-conception in the face of an other that it cannot itself eliminate. It can do so, on Hegel's account, only by relating to an object that *negates itself*. "On account of the independence of the object, [...] it can achieve satisfaction only when the object itself effects the negation within itself" [42, ¶175]. And since only entities with the cognitive capacity to deny, doubt, or otherwise revise their own self-understanding can self-negate, the only candidate for such an object is another self-consciousness. The concept of self-consciousness thus entails a plurality of self-consciousnesses. As Hegel puts it, "the truth of this certainty is really a double reflection, the duplication of self-consciousness" [42, ¶176].

The presence of another self-consciousness appears to resolve the problem generated by desire. Because the other is capable of negating itself, it presents itself not as an obstacle to self-consciousness but as a formal reflection of it. As Houlgate puts it, the other presents itself "as that which is nothing for itself, as that which is there not for its own sake but solely for the sake of the first self" [46, p88]. Hegel calls this complex relation between consciousnesses "recognition". Recognition allows self-consciousness to affirm itself as self-relating without abolishing the otherness to which it relates because, rather than eliminating the other, it encounters itself therein: "the other self-consciousness [...] gives it back again to itself" [42, ¶181]. Importantly, though, the other's self-negation does not abolish its independence. Indeed, the other can serve as an adequate reflection of the self-certainty of self-consciousness only in so far as it retains some degree of self-certainty in its own right. Though it sheds itself of all specific determinacies, it thus preserves a general sense of its own identity—or "it is *for itself* a genus" [42, ¶176]—and thus remains capable of freely reflecting the first self-consciousness: "Consciousness has for its object one which, of its own self, posits its otherness or difference as a nothingness, and in so doing is independent" [42, ¶176]. In this way, recognition appears capable of providing what desire could not—namely, self-certainty in the face of an irreducible other.

And yet, this solution also proves insufficient. Since the second self-consciousness is not merely an object of recognition but a self-consciousness in its own right, it possesses the same conceptual structure as the first. Accordingly, it too seeks affirmation of itself through recognition. Each self-consciousness therefore seeks recognition from the other while withholding it in return, since "it does not see the other as an essential being, but in the other sees its own self" [42, ¶182]. The result is a conflict that Hegel characterises as a life-and-death struggle.

Since recognition can be obtained only from another living self-consciousness, its very source disappears if the struggle results in death. In such a case, the conflict "does away with the truth which was supposed to issue from it" [42, ¶188]. The only logically instructive possibility is thus one in which both self-consciousnesses survive but one succeeds in subordinating the other. Hegel famously characterises this arrangement in terms of "lordship" and "bondage". The lord achieves the recognition she seeks, while the bondsman recognises the lord as independent and authoritative. *Prima facie*, this appears to resolve the recognitive problem. Self-consciousness seems finally to have secured the affirmation of itself that desire could not provide, and this by securing recognition from another self-consciousness. But this arrangement remains fundamentally unstable, and this because the recognition gained by the lord is not the reciprocal recognition of one self-consciousness by another. In Hegel's words: "The outcome is a recognition that is one-sided and unequal" [42, ¶191].

The difficulty here is that the lord achieves recognition only by depriving the recogniser of the very independence that made their recognition valuable in the first place. Recall that recognition originally appeared capable of resolving the problem generated by desire because it allowed self-consciousness to encounter itself reflected in *another independent self-consciousness*. Yet the relation of lordship transforms the bondsman into a dependent consciousness whose role is simply to affirm the lord's superiority: "the object in which the lord has achieved his lordship has in reality turned out to be something quite different from an independent consciousness" [42, ¶192]. The lord thus receives recognition, but not the kind of recognition originally sought. As Hegel puts it, the lord's "truth" lies in "the unessential consciousness" of the bondsman [42, ¶192]—that is, in a consciousness that cannot adequately mirror the lord's self-certainty. Lordship thus fails on its own terms: the very attempt to secure self-certainty through asymmetrical recognition undermines the conditions under which recognition can confer self-certainty at all.

The lesson Hegel draws from this failure is that recognition must be mutual. Self-consciousness can secure self-certainty neither by negating otherness (as in desire) nor by coercing recognition (as in lordship). Rather, it must encounter itself in another self-consciousness while simultaneously acknowledging that other as an independent self-consciousness in its own right. Hegel thus arrives at a relation of mutual recognition: in order to secure their self-certainty, multiple self-consciousnesses must "*recognize* themselves as *mutually recognizing* one another" [42, ¶184]. Only such a relation preserves the independence of both parties, thereby also allowing each to find itself reflected in the other. Mutual recognition thus resolves the internal inconsistencies that beset the logical structures of desire and lordship. Self-consciousness achieves certainty of itself not despite its relation to another self-consciousness, but through that relation. Mutual recognition, in other words, emerges as a necessary implication and constitutive condition of the very concept of self-consciousness.

Naturally, this reconstruction provides only a highly compressed survey of Hegel's account of mutual recognition. Nonetheless, it suffices to identify the mutually recognitive structure that emerges from his analysis and to consider the isomorphism between it and the implicit theoretical structure of QBism. Just as QBism implicitly claims that one can be a fully autonomous user of quantum mechanics only when one acknowledges other users as agents, so too does Hegel claim that one can be a self-conscious agent in a more general sense only by acknowledging others as equally free and self-conscious. If a user of quantum theory refuses to recognise another as a free agent, then they invite a situation not unlike Hegel's life-and-death struggle, wherein the commitments of each result in competing attempts to place one another in a state of suspended animation. We resolve such a situation for QBism only by respecting the other as a locus for action and experience, which is equivalent (according to our reconstruction in Section 2) to accepting limits on the scope of one's use of the theory—namely, limits defined by a personalist account of probability. Thus, QBists claim that the agency-negating threat of suspended animation is eliminated "once both friend and Wigner are understood as agents on an equal footing with regard to their individual uses of quantum theory" [1, p. 1859], just as Hegel claims that "I am only genuinely free when the other is also free and is recognized by me as free" [2, §431A].

We should be explicit about the scope and limits of this sort of recognitive account. Just as debate continues about the aims of Hegel's own argument, so too will questions arise about the specific implications of a recognitive account of agency for QBism. To give a clearer sense of these implications, consider Beiser's and the late Robert Stern's interpretations of Hegel. These are concerned with the adequacy of one's self-understanding and the extent of the realisation of one's freedom respectively: "the self knows itself to be a rational being only if it recognises the equal and independent reality of others, and only if the others recognise its own equal and independent reality" [47, p177]; "far from limiting and checking our freedom as it may at first seem, it is only by recognizing others as equal to ourselves that we can in fact realise that freedom" [48, p357]. Thus, the lesson for QBism is that to deny the consciousness of a friend is to fail to understand fully what it is to be a user of quantum mechanics, since such a denial is made on the basis of an overinflated understanding of one's use of the theory (i.e., as describing physical systems rather than organising one's probability assignments). Doubting the consciousness of a friend also hinders one's autonomous use of the theory, since one cannot *sincerely* doubt as much and still engage rationally (and so freely) in the essentially collaborative enterprise of scientific inquiry, one which presupposes the capacity of others to build experiments, make and revise state assignments, conduct measurements, report their results, and so on.

This latter point is inspired in particular by Stern's pragmatist interpretation of Hegel [49],[15] which—much like the phenomenological approach outlined in the previous section—undercuts scepticism about other minds by exposing its "empty and flawed view of epistemology" [48, p345]. Genuine scepticism is practically self-undermining because it seeks to establish knowledge without first engaging in the actual practice of acquiring it.[16] As Hegel himself puts it: "to want to have cognition *before* we have any is as absurd as the wise resolve of Scholasticus to learn to *swim before he ventured into the water*" [50, §10]. The point of a recognitive account of agency on this reading is thus not to prove the existence of other minds, but to undermine the very concern with that question, and in doing so to acknowledge what Stern calls "the sociality of freedom" [48, p359]. Just as one cannot learn to swim without entering the water, so too do we fail to understand the use, and indeed limit our own free and rational use, of quantum mechanics by neglecting the actual laboratory settings in which others are encountered as fellow agents.

Locating QBism within a broader tradition of mutual recognition allows us to identify and clarify its theoretical structure—that is, its implicit commitment to a mutually recognitive account of agency. It also grants the QBist access to novel conceptual resources and different orders of explanation. According to my reconstruction of QBism, mutually recognitive relations between users of quantum theory are a formal requirement for a coherent interpretation of its intersubjective application. For the phenomenologist, such relations are implicated in the structure of experience. On Hegel's account, by contrast, these relations obtain as a matter of conceptual or logical necessity. The very idea of a self-conscious agent just entails the recognition of other self-conscious agents. Hegel thus provides the strongest possible articulation of the conception of agency that QBism appears to entail: one in which agents are constitutively enmeshed in relations of mutual recognition with other agents.

## 5. Conclusion: mutual recognition in the philosophy of physics

In a recent overview of QBism, and referring specifically to the view's concepts of "agent" and "user of quantum mechanics", Fuchs remarks: "Boy, the phenomenologists might really help us here! The shortcomings will be quite apparent" [7, p. 82]. This article takes up that invitation. By reconstructing and disambiguating its response to Wigner's friend paradox, I argue that QBism commits itself implicitly to a recognitive conception of agency.

---

[15] I emphasise Stern's pragmatism partly because affinities between QBism and pragmatism have already been identified [7].
[16] Hence pragmatist C. S. Peirce's description of such scepticism as mere "paper doubt" [as cited in 49, p790].

Identifying this implication allows us to clarify what is at stake in QBist engagements with phenomenology—namely, a philosophically more robust conception of agency that might underpin the formal implications of QBism's handling of the intersubjective use of quantum theory. Importantly, we can extend this analysis by tracing the relevant phenomenological themes back to Hegel's account of mutual recognition. Doing so enriches our understanding of QBism as a novel development within a longstanding philosophical tradition while also affording potential new resources for the rigorous expression and justification of QBist ontology.

My claim is not necessarily that QBists should start carrying around a copy of the *Phenomenology of Spirit*. They are of course welcome to do so, and this essay shows how that text provides powerful resources for articulating the conception of agency implicated within their view.[17] More importantly, though, I argue that situating QBism within this broader tradition helps to clarify the status and prospects of that conception itself. QBism's mutually recognitive conception of agency is not as unorthodox as it may appear to be, nor need it be understood as a merely formal response to a technical puzzle. Rather, it belongs to a longstanding tradition in which relations to other agents are treated as constitutive of agency itself. Thus, the emerging QBist conception of agency enjoys a plurality of philosophical antecedents, motivations, and potential modes of justification. This essay has traced only one thread through that broader conceptual terrain.

More generally, this analysis illustrates the value of bringing contemporary philosophy of physics into dialogue with philosophical traditions not typically associated with it—the phenomenological and German idealist traditions in this case. The interpretative problems posed by quantum theory concern not only the nature of physical reality, but also the natures of agency, experience, and intersubjectivity. Thus, it would be surprising if traditions so centrally concerned with these topics proved to have nothing to contribute to their elucidation. Indeed, engagement with such traditions affords not only new conceptual resources, but also a degree of critical distance from contemporary assumptions (e.g., those structuring the quantum interpretation program [51, pp357–9,52]), allowing us to reflect anew on current practice. The more quantum theory continues to resist conceptualisation, the more reason it gives us to look beyond the methods and conceptual frameworks that have so far dominated discussion. The challenge may lie not only in revising our understanding of physical reality, but in reconsidering *how we seek* to understand it.

## Acknowledgments

[17] Assuming, of course, that one is willing to embrace the Hegelian way of doing philosophy.

I am grateful to James Read and Chris Fuchs for helpful discussions and comments on this work, and to the editors of this special issue and two anonymous reviewers for their careful and constructive feedback. I would also like to thank audiences at the New Approaches to the Foundations of Quantum Mechanics conference (Linköping, 27–29 November 2024); the Early Career Work in Progress Seminar at the University of Oxford (25 June 2025); the 21st Annual Conference of the Nordic Society for Phenomenology (University of Southern Denmark, 28–30 April 2025); and the Society for European Philosophy Annual Conference (King's College London, 7–9 July 2025) for valuable questions, comments, and discussion.